\documentclass{mem}
\usepackage{natbib}\usepackage{txfonts}\usepackage{balance}
\usepackage{graphicx}
\idline{75}{282}
\begin{document}
\def\teff{$T\rm_{eff }$}
\def\kms{$\mathrm {km s}^{-1}$}

\title{GPU-accelerated broadband analysis of multi-messenger light curves of GRBs}
\subtitle{}

\author{Maurice H.P.M.\,van Putten}

\institute{ Physics and Astronomy, Sejong University, 143-747 Seoul \email{mvp@sejong.ac.kr} }

\authorrunning{van Putten}

\titlerunning{GPU-analysis}

\abstract{
High-frequency multi-messenger observations provide a powerful probe of 
gamma-ray bursts (GRBs), pioneered by {\em BeppoSAX} in gamma-rays and LIGO-Virgo in 
gravitational waves. Identifying the central engines - magnetars or black holes - also promises to 
improve on GRBs as probes of cosmological evolution. THESEUS' design is ideally suited to
pursue both science objectives. Here, we present a general-purpose {\em graphics processor units} 
(GPU)-accelerated broadband search algorithm for long duration ascending and descending chirps,
post-merger or from core-collapse of massive stars, in electromagnetic and gravitational radiation. 
It implements butterfly filtering using banks of up to 8 million templates of 1 second duration at over 
one million correlations per second by heterogeneous computing using a dozen high-end GPUs. 
We demonstrate its application with the identification of broadband Kolmogorov spectra in long GRBs
and the long duration ascending chirp in the merger GW170817. The first shows a noticeable absence 
of a high frequency bump, otherwise expected from newly formed magnetar central engines. The 
second illustrates the need for deep searches to identify GRB central engines in descending chirps in 
gravitational waves, post-merger or from nearby energetic core-collapse supernovae. A future
catalogue of THESEUS' GRBs covering a broad range of redshifts may probe the nature of the 
cosmological vacuum and establish the de Sitter limit as a turning point in cosmological evolution.
}
\maketitle{}

\section{Introduction}

\begin{figure*}[t!]
\resizebox{\hsize}{!}{\includegraphics[clip=true]{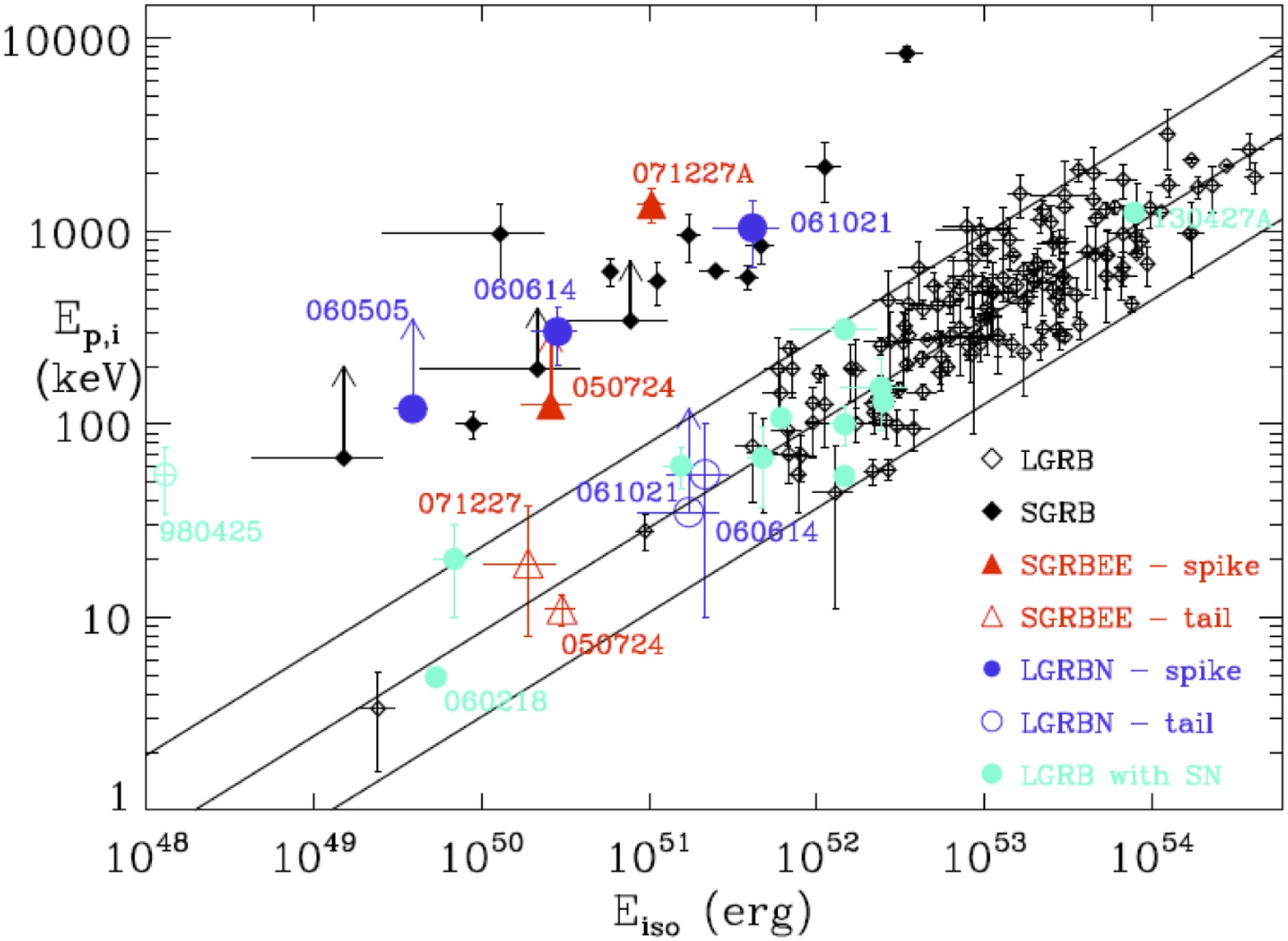}\includegraphics[scale=30]{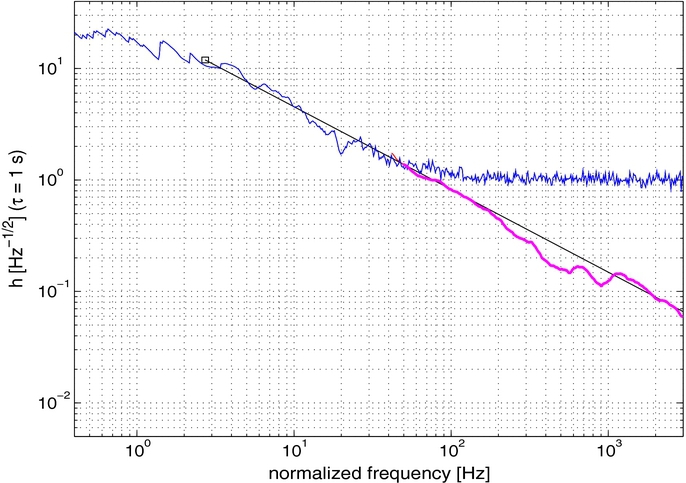} }
\caption{\footnotesize
(Left panel.) The $E_{p,i}-E_{iso}$ plane showing short GRBs with Extended Emission (SGRBEE) and long GRBs with no apparent association with supernovae (LGRBN)\citep{ama02,ama06}. Included are GRB-SNe 030329, 050525A, 081007,091127,100316D,101219B along the Amati-relation for normal long GRBs ($\pm2\sigma$). The sub-energetic GRB980425/SN1998bw is anomalous. The tails of GRBEEs 050724 and 060614 (also a LGRBN) (open triangles, red) falls well within the group of tails of normal LGRB (medium sized filled circles, green). The initial short-hard spike of SGRBEEs (solid triangles, red) falls into the separate group of SGRBs, in common with the initial pulse of LGRBNs (large size filled circle, blue).  (Reprinted from \citep{van14b}.)
(Right panel.) The broadband Kolmogorov spectrum (averaged over 42 spectra of long GRBs from {\em BeppoSAX}) extracted by butterfly filtering using a bank of 8.64 million templates show a featureless extension to over 2 kHz (comoving frame, purple). Noticeably absent is a bump at high frequency, that would otherwise be expected from magnetars newly formed with random orientations of magnetic moment and angular momentum.
(Reprinted from \citep{van14a}.)}
\label{figAm1}\label{figKol}
\end{figure*}

The nature of the central engine of gamma-ray bursts (GRBs) remains a key outstanding question,
currently probed indirectly by multi-wavelength observations in electromagnetic radiation \citep{pir04,nak07,zha16,zig17}. 
Based on total energetics and short-time scale variability \citep[e.g.][]{sar97,sar98}, 
their central engines are probably magnetars \citep[e.g.][]{met11} or
black holes \citep{woo93,eic00,woo06,woo10}. Since short GRBs (SGRBs) and SGRBs with Extended Emission (SGRBEEs) 
discovered by {\em Swift} derive from mergers, of neutron stars with another neutron star or black hole, their final 
remnants should be black holes. While the central engine to the initial short hard spike in SGRBs is uncertain, any Extended 
Emission {\em post-merger} lasting tens of seconds is likely associated with a black hole. Detailed spectral-energy relations 
for Extended Emission (SGREE) and the prompt emission of long GRBs (LGRBs) further show a remarkably
common Amati relation \citep{ama02} (Fig. \ref{figAm1}). Conceivably, therefore, Extended Emission
to SGRBs, if present, and long GRBs share a common inner engine: rotating black holes
producing extended emission in possibly both the electromagnetic and gravitational-wave spectrum over their 
lifetime of rapid spin \citep{van03,van12a}. 
\begin{figure*}[]
\resizebox{\hsize}{!}{\includegraphics[clip=true]{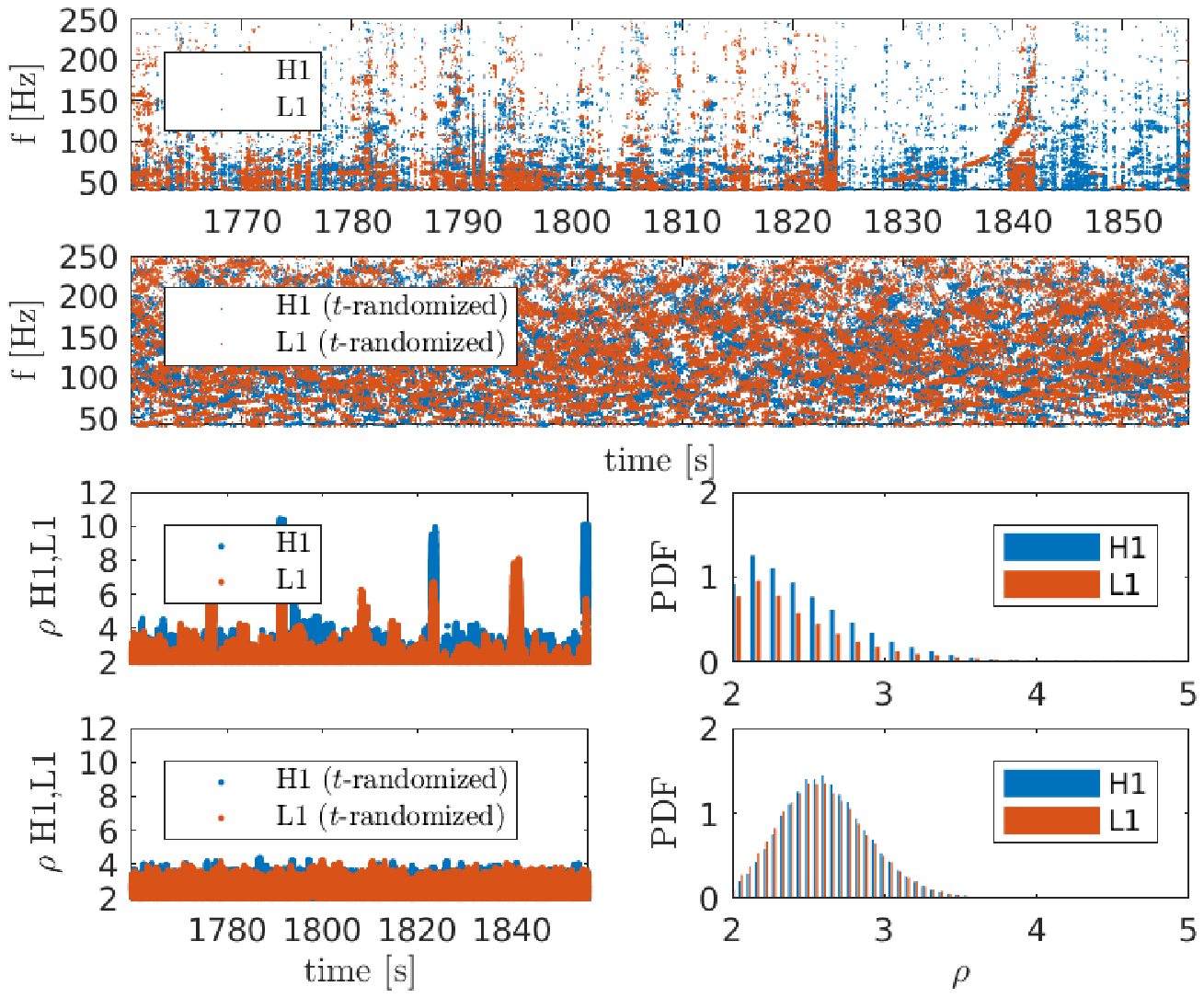}\includegraphics[scale=0.78]{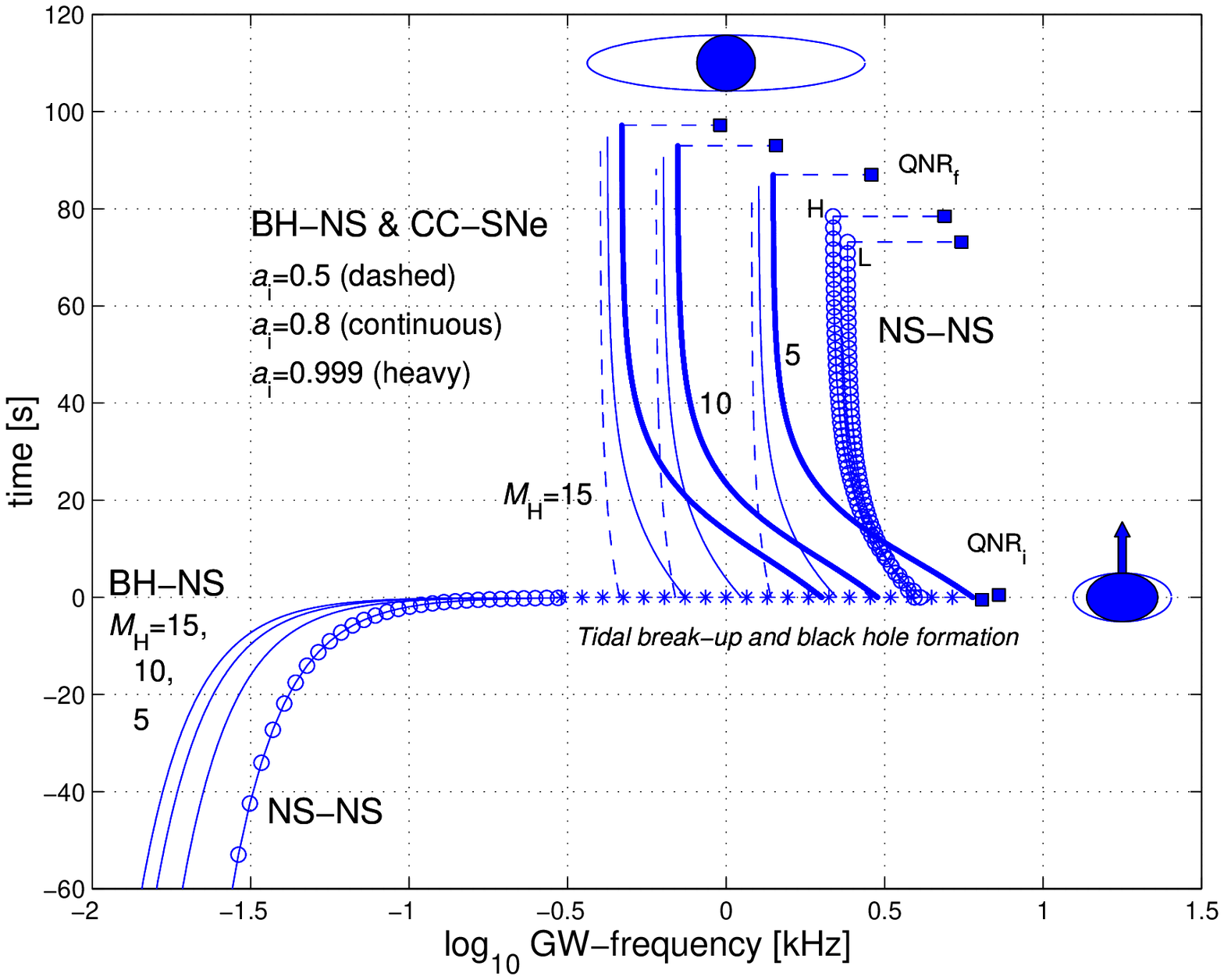}}
\caption{
\footnotesize
(Left panels.) The merger GW170817 produces an {\em ascending} chirp in gravitational waves, here shown as the output of
butterfly filtering of the LIGO H1 and L1 detectors, including control following time randomization (maximal entropy data).
The left-truncated skewed Gaussians of maxima of butterfly output are markedly different for the original data and control.
(Right panel.) Rotating black holes surrounded by a high density disk are a common outcome post-merger - of neutron stars
with another neutron star or companion black hole - and core-collapse of massive stars. Black holes thus formed may spin rapidly, 
in which case they may loose angular momentum to high density matter at the Inner Most Stable Circular Orbit (ISCO) leading to 
catalytic conversion of spin energy to gravitational radiation \citep{van03}. In this process, the outcome is a {\em descending} chirp 
in gravitational waves for the lifetime of rapid spin. For GW170817 type events, the asymptotic result is tightly constraint by the 
narrow distribution of neutron star masses and spin \citep{bai08}, here indicated by $H$ and $L$ for high- and low-mass neutron stars. 
Absent a remnant stellar envelope, such binary merger creates a naked inner engine, whose magnetic winds may produce an 
observable radio burst. (Reprinted from \citep{van09}.)} 
\label{figGW17}
\end{figure*}
To rigorously identify the central engine to GRBs by gravitational-wave detection \citep{cut02}, 
we propose a focus on high-frequency broadband 
analysis of contemporaneous emission in electromagnetic and gravitational radiation. While both
magnetars and black holes - interacting with high density matter formed post mergers and core-collapse 
of massive stars - can produce extended emission in the process of spin down, their spectral
properties should be quite different. In the electromagnetic spectrum, magnetars newly formed 
are expected to produce a broadband bump around 2 kHz (co-moving frequency) by generic
mis-alignment of magnetic moment and angular momentum, whereas this would be absent
for any high-energy emission from black hole outflows by inherent alignment of the same by
Carter's theorem \citep{car68}. No such bump is seen in a broadband analysis of light curves of a sample of bright long
GRBs from the {\em BeppoSAX} catalogue (Fig. \ref{figKol}). 

Furthermore, any gravitational wave emission from a magnetar or a black hole formed in above mentioned
extreme transient events should be a {\em descending chirp}, 
as angular momentum carried off in gravitational radiation of a single object causes spin down, as opposed to spin
up for binary systems. For magnetars, such descending chirps formally can reach zero frequency, whereas 
for black holes of mass $M$, the late time frequency - from non-axisymmetric quadrupole mass motion about their 
Inner Most Stable Circular Orbit (ISCO, \cite{ker63,bar72}) - reaches a late time plateau \citep{van11} (Fig. \ref{figGW17})
\begin{eqnarray}
f_{GW} = (600-700)\, \mbox{Hz} \left( \frac{M}{10M_\odot}\right)^{-1}
\label{EQN_fGW}
\end{eqnarray}
at a stationary point when the angular velocities of the black hole and matter at the ISCO are equal.
In (\ref{EQN_fGW}), the frequency uncertainty reflects variations in initial spin.
For GW170817 like events with $M\simeq 3M_\odot$, $f_{GW}$ would gradually settle down to a 
late time frequency of about 2000 Hz. This illustrates the need for high frequency multi-messenger probes 
 to unambiguously identify GRB central engines. 
Establishing rotating black holes as central engines to GRBs and energetic core-collapse
supernovae would give a first probe of relativistic frame dragging interactions with high
density matter. Such measurement of gravitational interactions with matter is 
a limit of {\em strong} gravitation, representative for the theory of general relativity. 

Extreme transient events such as GW170817 \citep{abb17} and GRBs also have the potential to probe
cosmological evolution \citep{ama12,ama13,ama16} completely independent of the use of Type Ia supernovae and Cepheids. 
Already, GW170817 and its accompanying GRB170817A give an interesting estimate of the Hubble 
parameter $H_0$ \citep{gui17}, completely independent of existing $H_0$ estimates.
In general, cosmological evolution represents {\em weak} gravitational interactions about the de Sitter scale 
of acceleration 
\begin{eqnarray}
a_{dS}=cH, 
\end{eqnarray}
set by the Hubble parameter $H=H(z)$ as a function of redshift $z$ and velocity of light $c$. 

\begin{figure*}[]
\resizebox{\hsize}{!}{\includegraphics[scale=0.42]{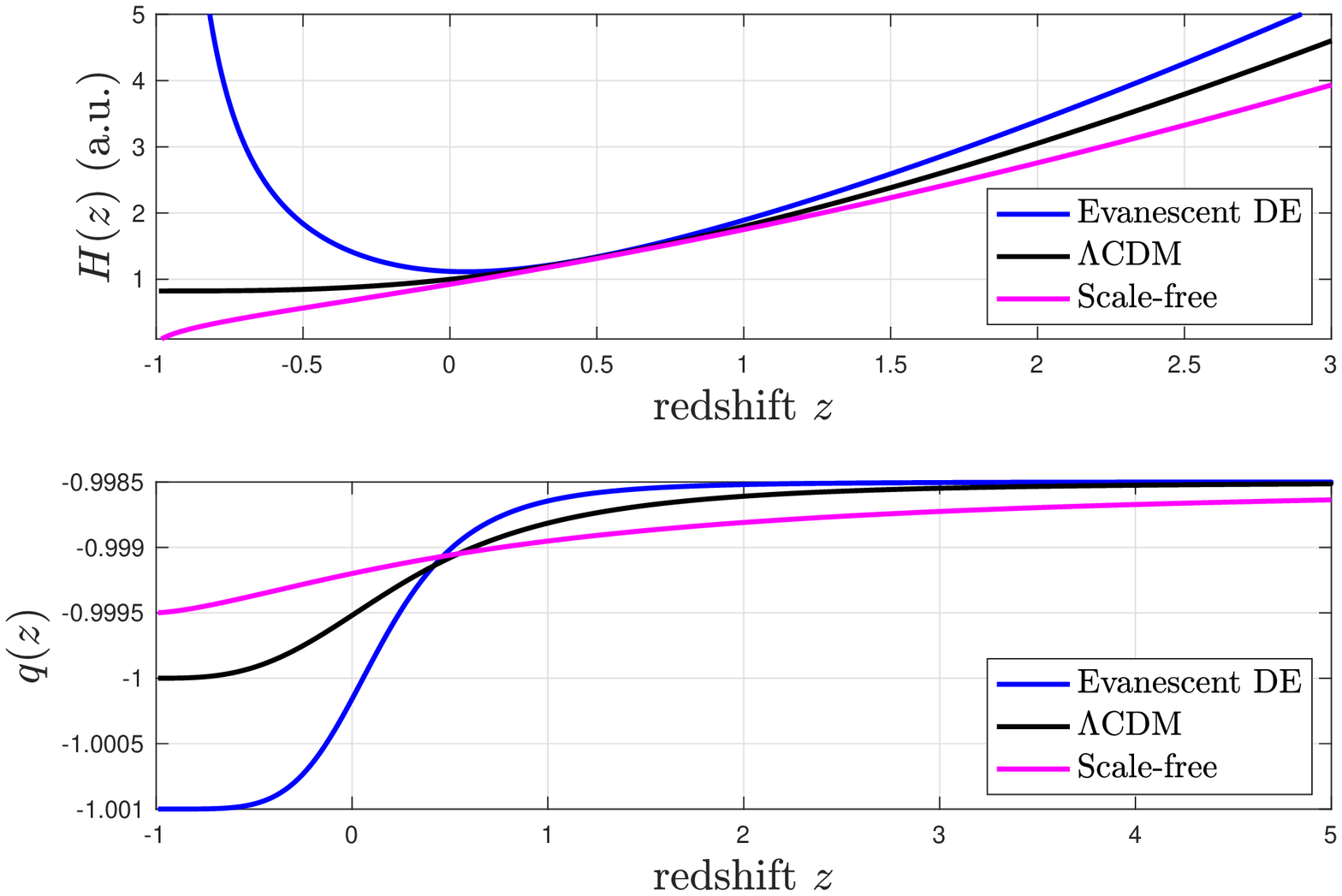}\includegraphics[scale=0.4]{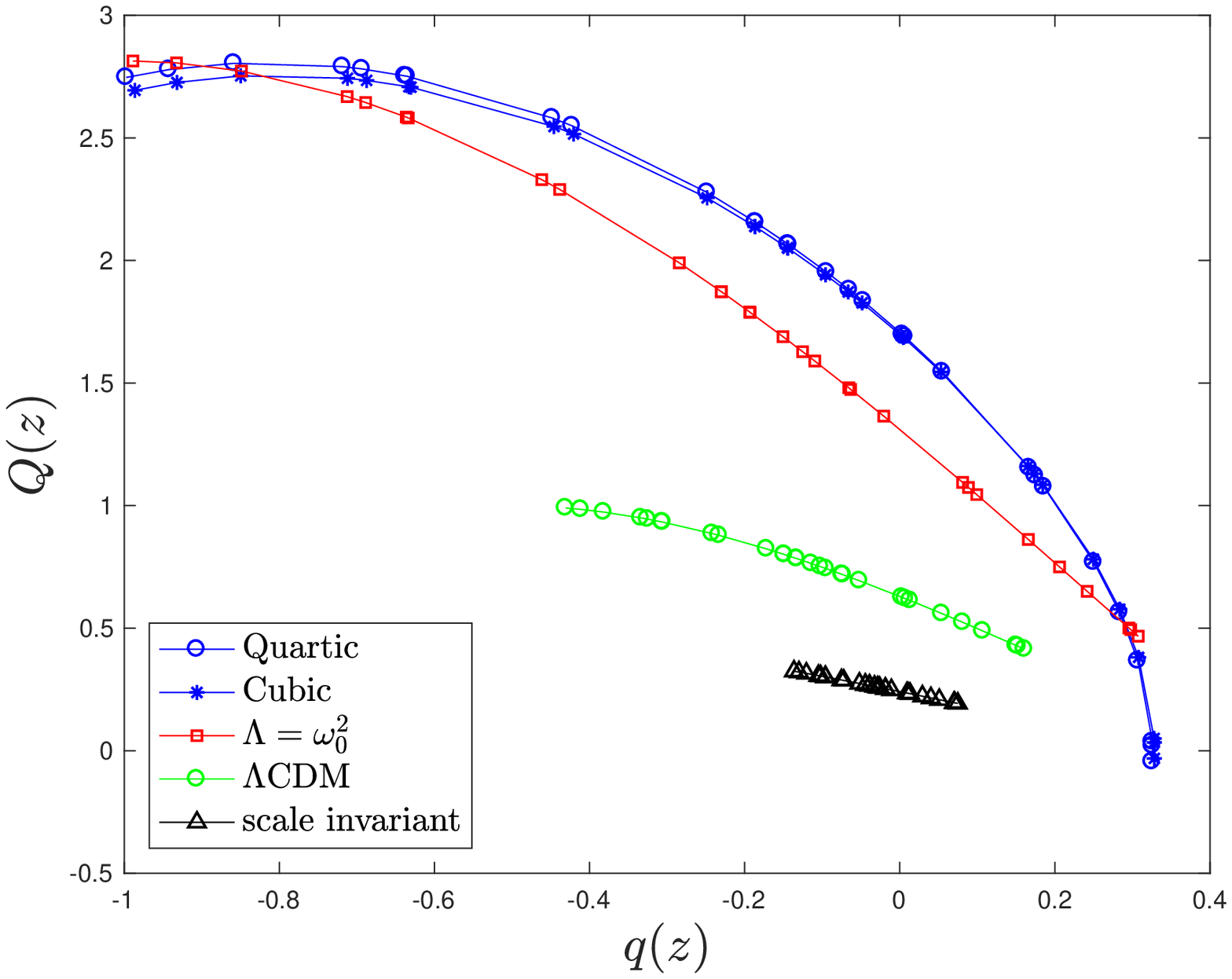}}
\caption{
\footnotesize
(Left panel.) Evolution of the Hubble parameter $H(z)$ as a function of redshift 
according to evanescent dark energy, $\Lambda$CDM and a scale-free cosmology.
(Right panel.) Confrontation of the three scenarios with heterogeneous $H(z)$ data 
over $0<z<1$ by nonlinear model regression, showing agreement of
evanescent dark energy but not $\Lambda$CDM with model-independent fits by cubic and quartic polynomials.
This result indicates that the de Sitter state $(q=-1, H^\prime(z)=0)$ is a turning point, rather than an asymptotically
stable state in $\Lambda$CDM. Similar analysis may be pursued by a large sample
of GRBs covering an extended range of redshift. (Extended from \citep{van17c}.)}
\label{figH}
\end{figure*}

Conceivably, weak gravitation takes us away from the geometric optics limit 
described by classical general relativity. In particular, evanescent dark energy and dark matter is expected
from super-horizon scale fluctuations leaking in through the cosmological horizon \citep{van17c} (Fig. \ref{figH}).
According to the deceleration parameter 
\begin{eqnarray}
q(z)=-1+(1+z)H^{-1}(z)H^\prime(z),
\label{EQN_q}
\end{eqnarray}
the de Sitter limit $(q=-1, H^\prime(z)=0)$ is hereby a {\em turning point} ($H^\prime(z)=0$) in cosmological evolution,
showing a cosmological vacuum beyond the classical limit envisioned in general relativity. 

The proposed {\em Transient High Energy Sky and Early Universe Surveyor} (THESEUS) mission 
\citep{ama17} is designed to detect high-resolution light curves from GRBs over a broad range of 
redshift, which is ideally suited to pursue the above science in {\em strong and weak gravitational
interactions} in central engines and, respectively, cosmological evolution. 

For deep searches in broadband light curves of gamma-rays and gravitational radiation, we recently
developed a new pipeline of butterfly filtering (\S2), accelerated by {\em graphics processor units} (GPUs). 
\S3 presents the need for accurate probes of $H(z)$ and $q(z)$ over an 
extended range of redshift $z$. We summarise these science objectives in \S4.

\section{Deep searches by butterfly filtering}

\subsection{Butterfly filtering} 

\begin{figure*}[t!]
\resizebox{\hsize}{!}{ \includegraphics[clip=true]{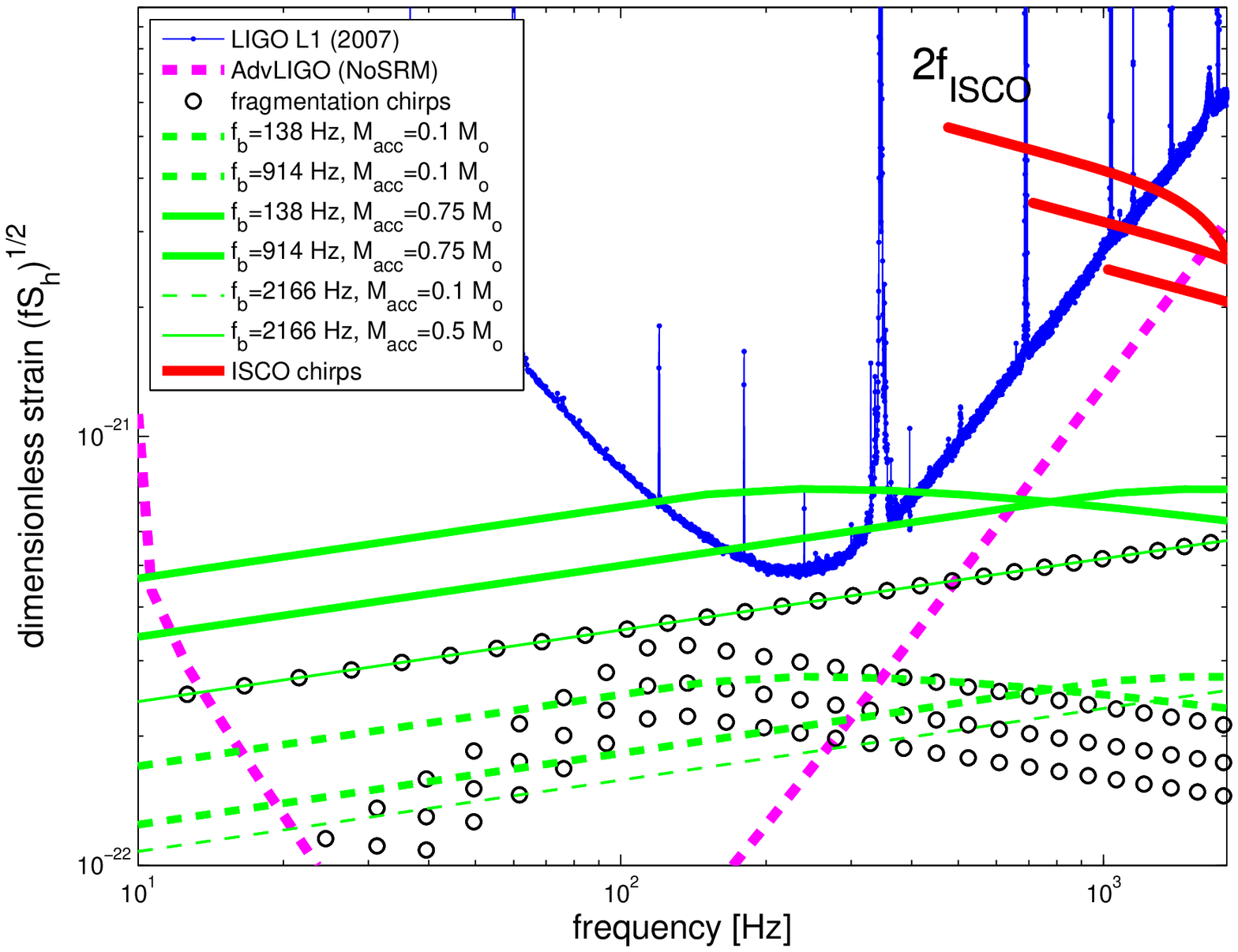}  \includegraphics[scale=1.5]{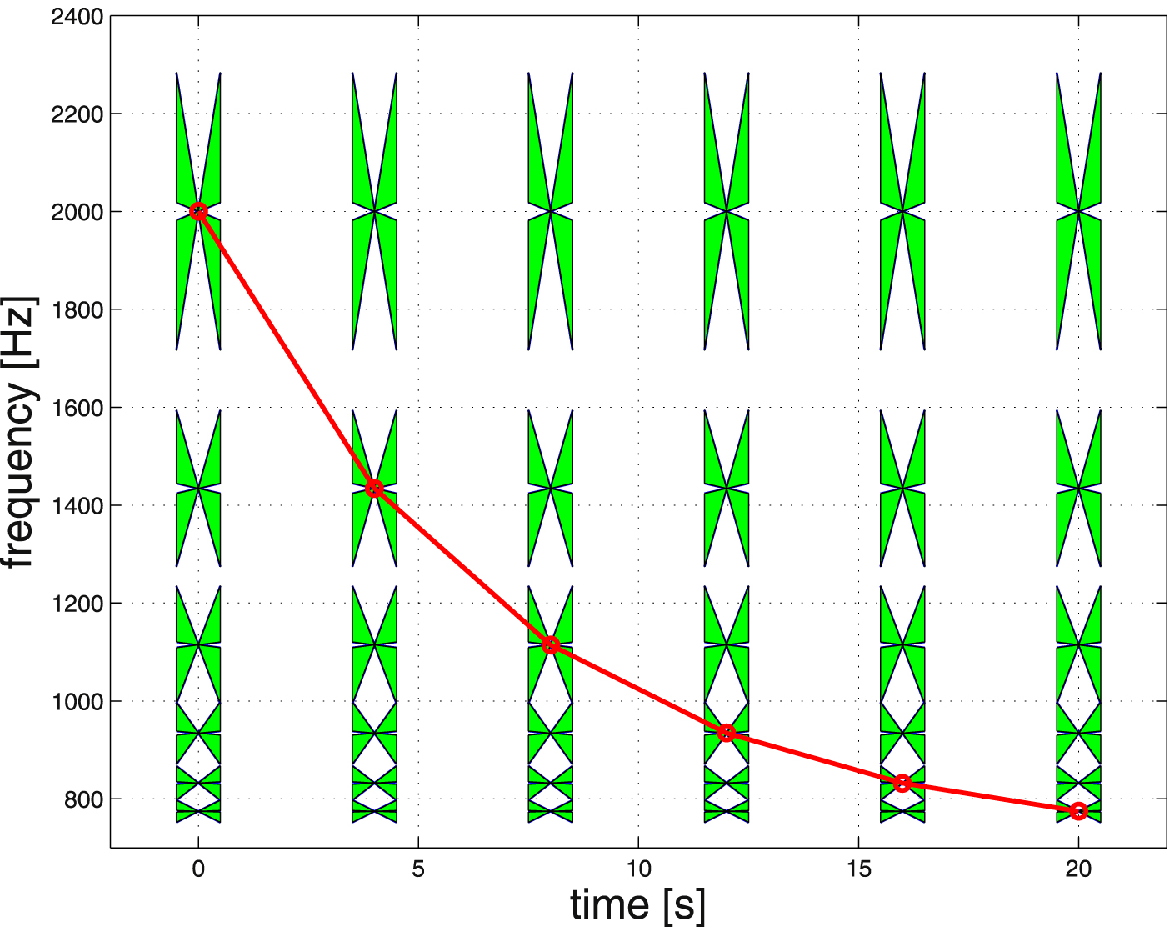} }
\caption{\footnotesize
(Left panel.) Characteristic strain $h_{char}(f)$ of quadrupole gravitational radiation from accretion flows around rotating black holes formed in core-collapse of massive stars at $D=100$ Mpc. The vertical distance to the dimensionless strain detector noise $h_n=\sqrt{fS_h}$ represents the maximal attainable S/N ratio obtainable by filtering. Model curves shown are broadband emission from non-axisymmetric accretion flows (green), fragmentation chirps of \citep{pir07} (circles, $\sigma_f=0.1$, $f_{e} = 120$ Hz) and ISCO waves induced by feedback from a central black hole (red). The curves shown refer to a black hole mass $M=10M_\odot$ (black), $M=7,10$ and $15M_\odot$ (green, red).
(Right panel.) Butterfly filtering is a bandpass filter of trajectories with finite slope $0<\delta \le df(t)/dt$ in frequency $f(t)$, suppresses signals with essentially constant frequencies, by matched filtering against a bank of a large number of templates of intermediate duration
covering a broad bandwidth in frequency. (Reprinted from \citep{lev15,van16}.) }
\label{figBF}
\end{figure*}

In powering extreme transient events - SGRBEEs, LGRBs and core-collapse supernovae - central engines exhaust their
reservoir in angular momentum \citep[for an early discussion in the context of core-collapse supernovae, see][]{bis70}
over tens of seconds. Their output in electromagnetic and gravitational radiation
will be unsteady and broadband with secular evolution in frequency, modulated by non-axisymmetric accretion flows 
(Fig. \ref{figBF}). For this reason, deep searches for their energetic output should be focused on frequencies that
gradually vary in time, i.e., long duration ascending or descending chirps as opposed to constant frequencies. 
This may be approached by butterfly filtering, that essentially side-steps ordinary Fourier analysis \citep{van14a,van16}. 
The resulting {\em chirp-based} spectrograms may reveal trajectories of frequency $f(t)$ with finite slope $df(t)/dt$,
deteced by matched filtering using a large bank of templates, that are superpositions of ascending and descending chirps 
of intermediate duration, e.g., on the order of 1 s - an educated guess of potential phase-coherence on intermediate time scales.

Fig. \ref{figKol} shows a demonstration in the identification of a broadband Kolmogorov spectrum in light curves of long
GRBs at an on-average 1.26 photon per 0.5 ms bin from the {\em BeppoSAX} catalogue, using a bank of 8.64 
million templates \citep{van14a}.

\subsection{GPU acceleration} 

Fig. \ref{figGW17} shows an application to the gravitational strain data of the LIGO H1 and L1 output during GW170817, using a 
similar bank of 4 million templates. The ascending chirp of the binary neutron star merger identified should be compared 
with our control, produced by the same analysis following time-randomisation of the H1 and L1 data. 
(Time-randomisation produces data with maximal entropy, keeping histograms the same.)

\begin{figure*}[]
\resizebox{\hsize}{!}{\includegraphics[clip=true]{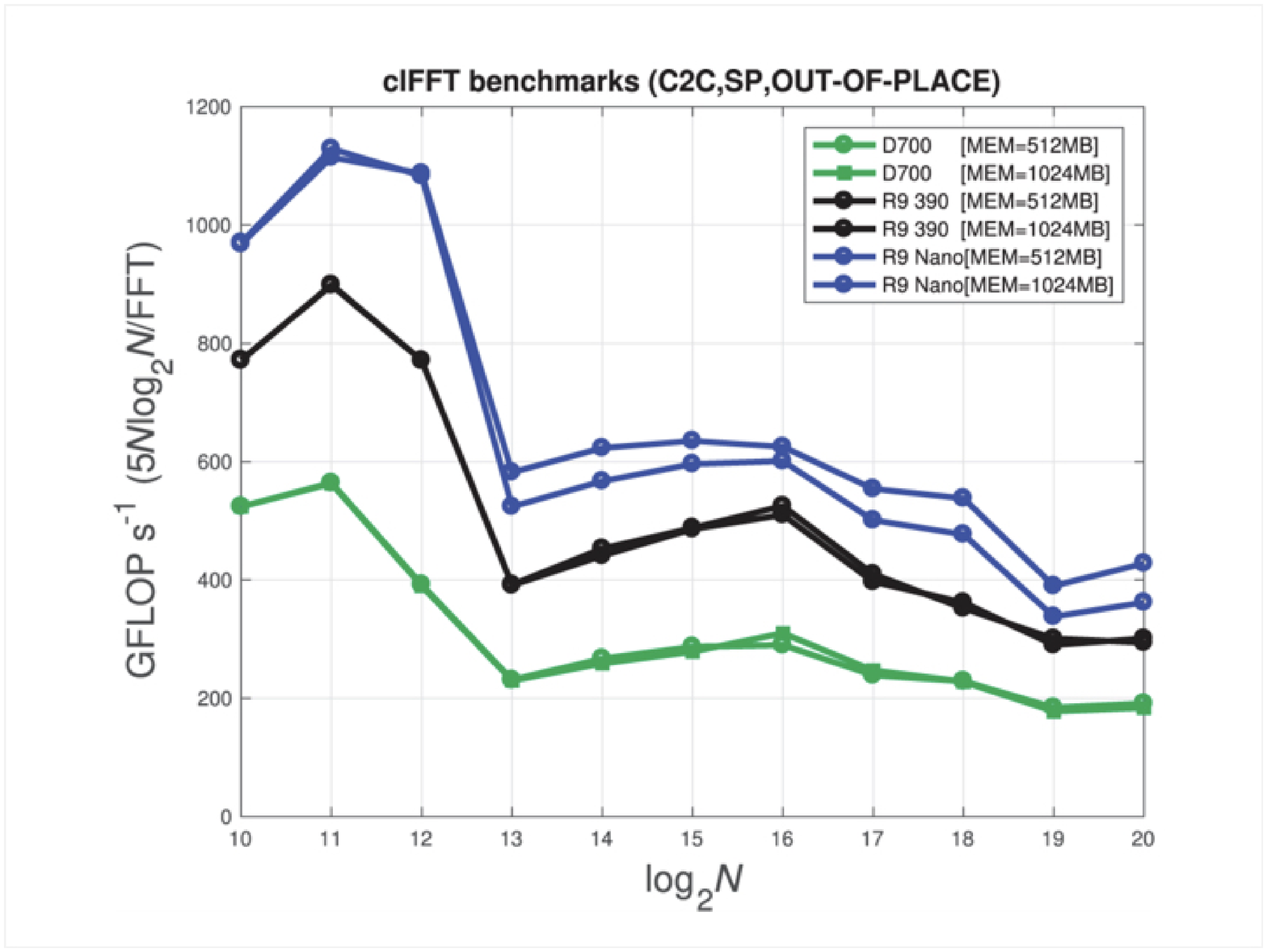}\includegraphics[clip=true]{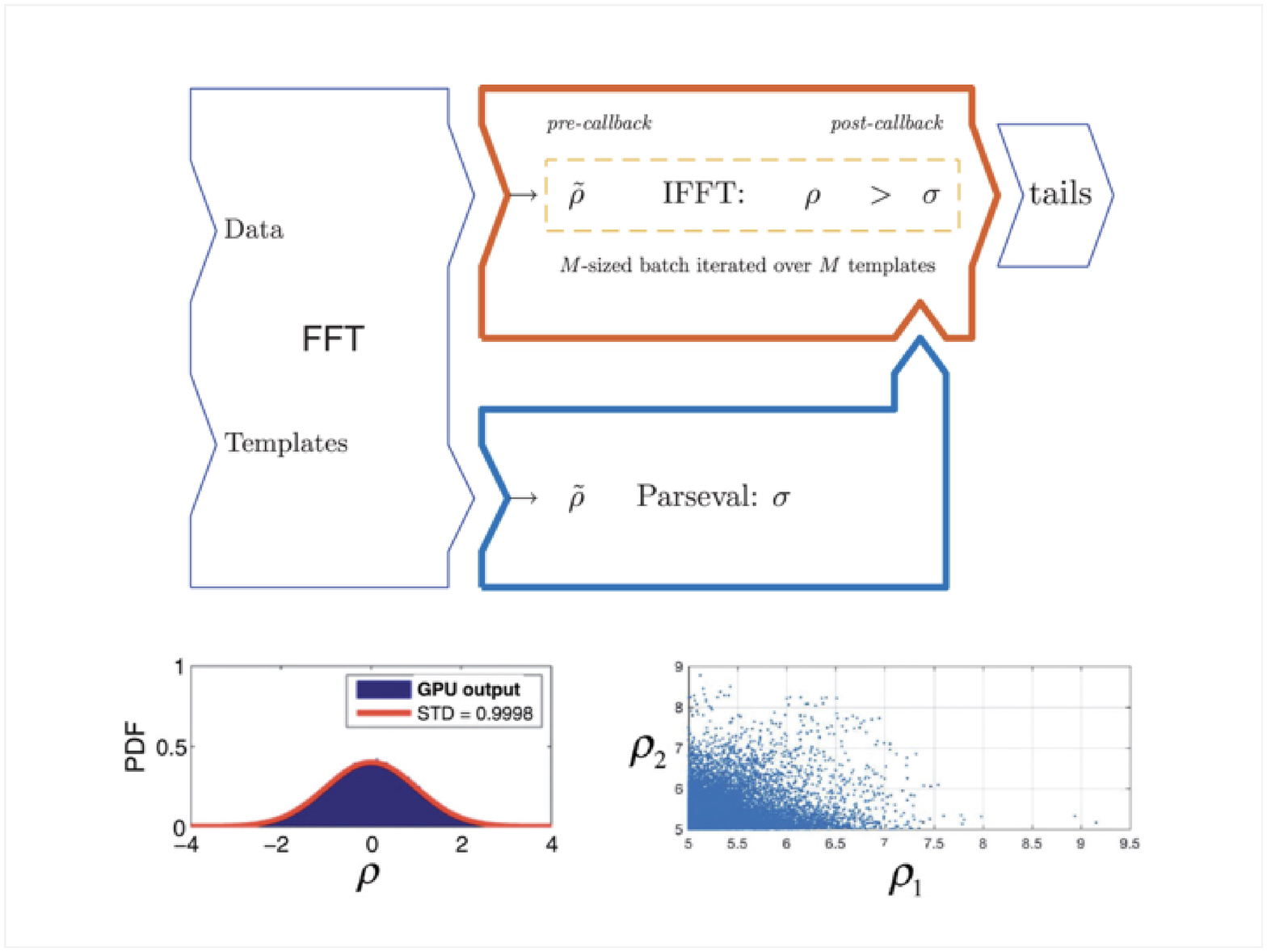}}
\caption{
\footnotesize
GPU benchmark results on clFFT$_N$ under OpenCL for complex-to-complex (C2C) single precision (SP) with out-of-place allocation in Global Memory for various high-end GPUs of Advanced Micro Devices (AMD). For data-segments of 16 seconds ($N=2^{16}$ at a sampling rate of 4096 s$^{-1}$), the R9 Nano with High Bandwidth Memory 2 (HBM2) enables about 100,000 transforms per second. Efficient GPU-acceleration obtains by circumventing limitations of the PCI by communicating only tails $\rho > \kappa\sigma$ back to the CPU from Global Memory in the GPU, where
$\kappa>1$ defines the search depth and $\sigma$ is the standard deviation of the matched filtered output $\rho$, computed in 
a predictor step by Parseval's Theorem (also off-loaded to the GPU). (Reprinted from \citep{van17b}.)}
\label{figBE}
\end{figure*}

\begin{figure*}[]
\resizebox{\hsize}{!}{\includegraphics[clip=true]{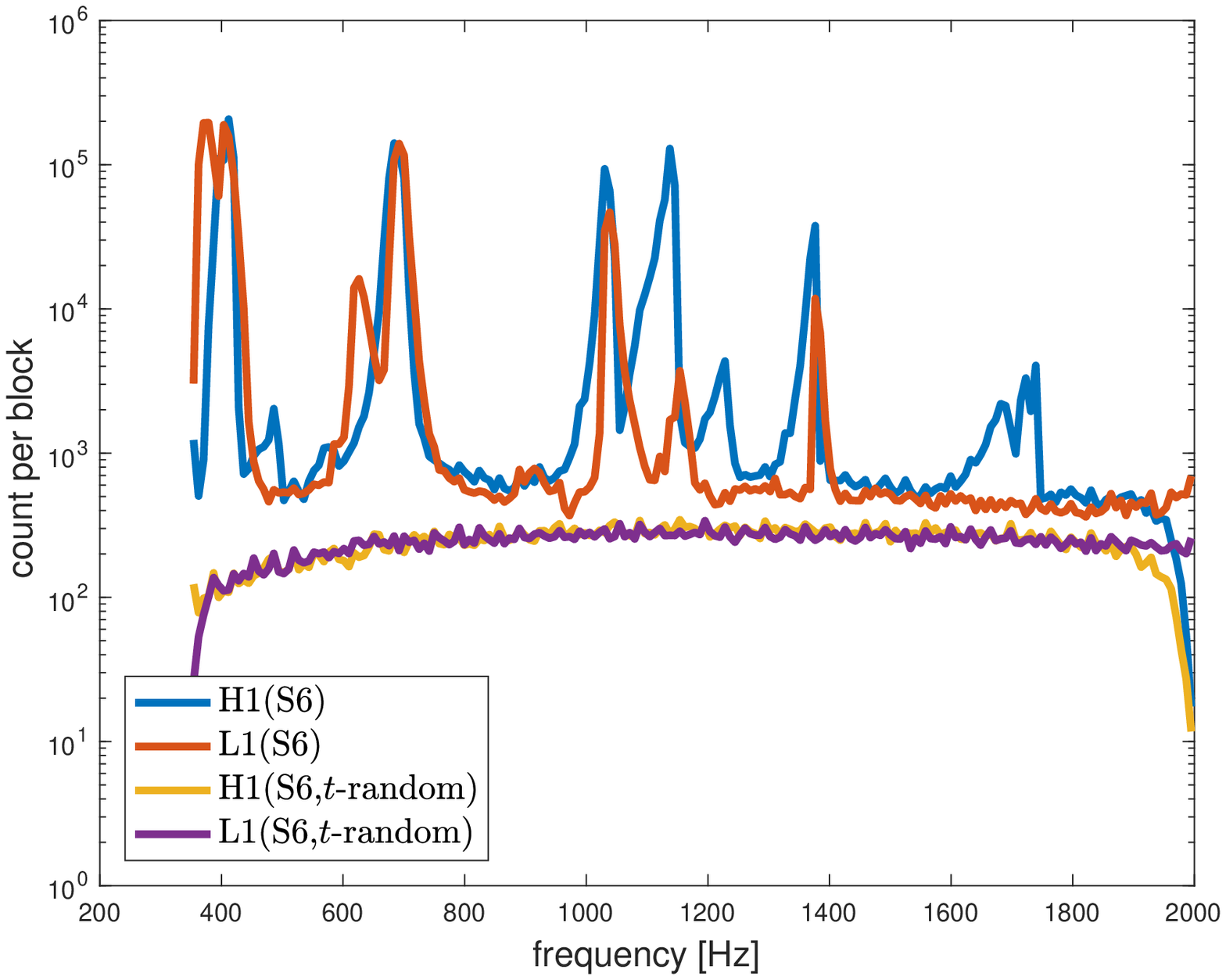}\includegraphics[clip=true]{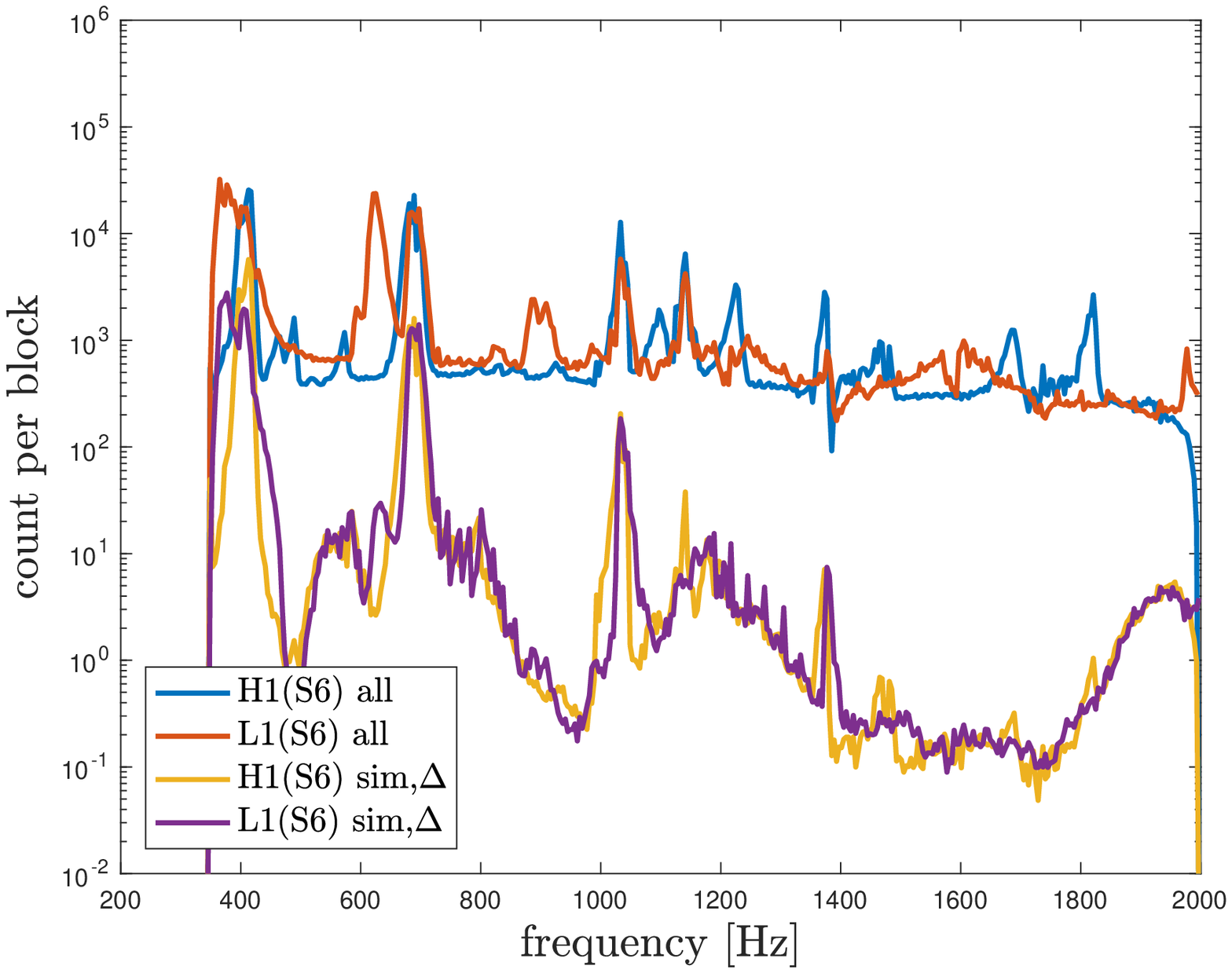}}
\caption{\footnotesize
(Left) Pseudo-spectra of simultaneous $\rho > \kappa\sigma$ in H1 (red) and L1 (blue), averaged over four blocks of S6 H1$\wedge$ L1, using a bank of 8M templates, along with control using maximal entropy (time-randomized) data. (Right) Pseudo-spectra as an average over all 288 blocks of S6 H1$\wedge$L1, using a bank of 0.5M templates, of H1 and L1 by independent counts and by simultaneous counts with frequency pairs within  $\Delta f$ = 50 Hz. (Reprinted from \citep{van17b}.)}
\label{figBFRE}
\end{figure*}

Fig. \ref{figGW17} is computed by heterogeneous computing accelerated by high-end GPUs \citep{van17b},
exploiting their high-performance processing of the Fast Fourier Transform (FFT, Fig. \ref{figBE})
and by circumventing bandwidth limitations of the {\em Peripheral Component Interface} (PCI) between 
GPU and the Central Processing Unit (CPU) using Parseval's Theorem. In this algorithm,
only tails of butterfly output $\rho$ exceeding $\kappa \sigma$, where $\kappa$ parameterises the
depth of the search ($\kappa>1$), are communicated to the CPU. Under the {\em Open Compute Language} (OpenCL), 
this algorithm realises over one million correlations per second on a platform with 12 {\em Advanced Micro Devices} 
(AMD) Fiji chips with {\em High Bandwidth Memory 2} (HBM2) at an overall efficiency about 65\%, 
normalized to clFFT GPU-performance. It enables a complete analysis of LIGO S6 by butterfly
filtering against a bank of up to 8 million templates within a one-year compute time (Fig. \ref{figBFRE}, 
comprising about $10^{20}$ floating point operations). 

\section{Probing the cosmological vacuum}

High resolution measurements of the Hubble parameter $H(z)$ and the associated deceleration parameter
(\ref{EQN_q}) offer a detailed probe of the cosmological distributions of dark energy \citep{rie98,per99}.
Late time evolution is particularly sensitive to the nature of dark energy (Fig. \ref{figH}). 
While evolution in $\Lambda$CDM - with a static dark energy - is relatively stiff ($H^\prime(0)\simeq 0.5H_0>0$),
evolution is relatively fast with evanescent dark energy derived from super-horizon scale fluctuations $(H^\prime(0)\simeq0$).
While the de Sitter state of cosmology is {\em assumed} to be the stable endpoint of cosmological evolution in $\Lambda$CDM,
it is a turning point in evolution by \citep{van15,van17c}
\begin{eqnarray}
\Lambda=\omega_0^2
\label{EQN_La}
\end{eqnarray}
from {\em evanescent fluctuations} (off-shell) at frequencies below the fundamental frequency
\begin{eqnarray}
\omega_0 = \sqrt{1-q}H
\label{EQN_ev1}
\end{eqnarray} 
of the cosmological horizon as an apparent horizon surface 
\citep{pen65,bre88,yor89,wal91,coo92,coo00,tho07}. With finite surface gravity away from the
radiation dominated regime \citep{kod80,hay98,hay99,bak00,cai05}, it gives rise to a stress
energy tensor of the cosmological vacuum with nonzero trace, that is, a cosmological distriubution
of evanescent dark energy and dark matter \citep{van18}.

The $qQ$-diagram ($Q(z)=dq(z)/dz)$, Fig. \ref{figH}) shows a confrontation of these model alternatives 
$H(z)=\sqrt{1+\omega_m(6z+12z^2+12z^3+6z^4+(6/5)z^5)}/(1+z)$ and, respectively,
$H(z)=\sqrt{1-\omega_m+\omega_m(1+z)^3}$, where $\omega_m$ denotes the cosmological density
of matter (baryonic and cold) today, with tabulated data on $H(z)$ over an extended range of redshift 
of \cite{far17}.
Included is further a confrontation of recently proposed scale-free cosmologies \citep{mae17,jes17}.
Agreement is found for evanescent dark energy with model-independent fits by cubic and quartic polynomials 
using nonlinear model regression. The latter rule out $\Lambda$CDM at a level of confidence of $2.7\sigma$ \citep{van17c}.
The resulting expectation for the Hubble parameter $H_0=H(0)$ is
\begin{eqnarray}
H_0 = 74.9\pm2.6 \mbox{km}\,\mbox{s}^{-1} \mbox{Mpc}^{-1},
\label{EQN_H0}
\end{eqnarray}
consistent with \cite{rie16,and17} and $H_0$ from GW170817 \citep{gui17}. 

In Fig. \ref{figH}, {\em gap} in $Q_0=Q(0)$, i.e., $Q_0>2.5$ for (\ref{EQN_La}) and $Q_0\lesssim1$ for $\Lambda$CDM,
represents a reformulation of the $H_0$ tension problem \citep{fre17}. Since this gap is of order unity, we expect that 
future samples of GW170817 type events and cosmological samples of GRBs from THESEUS over a broad range in 
redshifts will resolve this to better than the present $2.7\sigma$ \citep{van17c}. 


\section{Conclusions}

High frequency multi-messenger probes of light curves of GRBs will be instrumental in
identifying their central engine, pertinent to resolving their physical origin and improving
their potential as probes of cosmological evolution. Already, electromagnetic light curves 
increasingly point to black hole central engines rather than magnetars, at least for the soft
extended emission common to SGRBEEs and normal LGRBs. Rigorous confirmation
is expected from simultaneous detection of descending chirps in gravitational radiation,
that may be identified in GW170817 like events in the near future or by detection of
central engines to energetic core-collapse supernovae. 

THESEUS' design promises a significant advance to these science objectives, possibly in 
combination with an extended sample of GW170817 type events by LIGO-Virgo and
KAGRA \citep{aku18}, in seeking answers to the questions: 
{\em What happened post-merger in GW170817?} Was the central engine to GRB170817A 
a magnetar or black hole? {\em What is the nature of the cosmological vacuum?}
Is dark energy evanescent or constant, i.e., is the de Sitter state a turning point or an
endpoint in cosmological evolution? 

\begin{acknowledgements}
The author gratefully acknowledges stimulated discussions with 
L. Amati, F. Frontera, M. Della Valle, A.J. Weinstein and J.B. Kanner and support from NRF of Korea under 2015R1D1A1A01059793 and 2016R1A5A1013277. LIGO S6 and O2 data are from LOSC provided by LIGO and the LSC. LIGO is funded by the U.S. NSF. Additional support is from MEXT, JSPS Leading-edge Research Infrastructure Program, JSPS Grant-in-Aid for Specially Promoted Research 26000005, MEXT Grant-in-Aid for Scientific Research on Innovative Areas 24103005, JSPS Core-to-Core Program, Advanced Research Networks jointly with the ICRRR, Tokyo.
\end{acknowledgements}

\bibliographystyle{aa}

\begin{thebibliography}{}
\bibitem[Abbott et al.(2017a)]{abb17} Abbott, B.P., Abbott, R., Abbott, T.D., et al., 2017, Phys. Rev. Lett., 119, 161101
\bibitem[Akutsu et al.(2018)]{aku18} Akutsu, T., Ando, M., Araki, S., Araya, A., Arima, T., et al., 2018, PTEP, 013F01
\bibitem[Amati et al.(2002)]{ama02} Amati, L, Frontera, F, Tavani, M, in 't Zand, J.J.M., Antonelli, A, et al., 2002. A\&A, 390, 89
\bibitem[Amati(2012)]{ama12} Amati, L., 2012, IJMP Conf. Ser., 12, 19
\bibitem[Amati \& Della Valle(2013)]{ama13} Amati, L., \& Della Valle, M., 2016, IJMP-D, 22, 1330028
\bibitem[Amati \& Della Valle(2016)]{ama16} Amati, L., \& Della Valle, M., 2016, A\&A Tr., 29, 193
\bibitem[Amati et al.(2017)]{ama17} Amati, L., O'Brian, D., Goetz, D., Bozzo, E., Tenzer, F., et al., 2017, arXiv:1710.04638

\bibitem[Anderson \& Riess(2017)]{and17} Anderson, R.I., \& Riess, A.G., 2017, arXiv:  1712.01065v1
\bibitem[Baiotti et al.(2008)]{bai08} Baiotti, L., Giacomazzo, B., \& Rezzolla, L., 2008, Phys. Rev. D, 78, 084033
\bibitem[Bak \& Rey(2000)]{bak00} Bak, D., \& Rey, S.-J., 2000, Class. Quant. Grav., 2000, L83
\bibitem[Bardeen et al.(1972)]{bar72} Bardeen, J.M., Press, W.H., \& Teukolsky, S.A., 1972, Phys. Rev. D, 178, 347 
\bibitem[Bisnovatyi-Kogan(1970)]{bis70}Bisnovatyi-Kogan, G. S., 1970, Astron. Zh., 47, 813
\bibitem[Brewin(1988)]{bre88} Brewin, L., 1988, Phys. Rev. D, 38, 3020
\bibitem[Cai \& Kim(2005)]{cai05} Cai, R.-G., \& Kim, S.P., 2005, JHEP, 2, 50 
\bibitem[Carter(1968)]{car68}Carter, B., 1968, Phys. Rev., 174, 1559
\bibitem[Cook \& Abrahams(1992)]{coo92} Cook, G.B., \& Abrahams, A.M., 1992, Phys. Rev. D, 46, 702
\bibitem[Cook(2000)]{coo00} Cook, G.B., 2000, Liv. Rev. Rel., 3, 5
\bibitem[Cutler \& Thorne(2002)]{cut02}Cutler, C., \& Thorne, K.S., 2002, in Proc. GR16, Durban, South Afrika
\bibitem[Eichler \& Levinson(2000)]{eic00} Eichler, D. \& Levinson,A. 2000, ApJ, 529, 146 
\bibitem[Farooq et al.(2017)]{far17} Farooq, O., Madiyar, F.R., Crandall, S., \& Ratra, B., 2017, {\it ApJ}, {\bf 835}, 26
\bibitem[Freedman(2017)]{fre17} Freedman, W. L. 2017, NatAs, 1, 0121
\bibitem[Guidorzi et al.(2017)]{gui17} Guidorzi, C., Margutti, R., Brout, D., Scoling, D., Fong, W., et al., 2017, ApJ, 851, L36
\bibitem[Hayward(1998)]{hay98} Hayward, S.A., 1998, Class. Quant. Grav., 15, 3147

\bibitem[Hayward et al.(1999)]{hay99} Hayward, S.A., Mukohyama, S., \& Ashworth, M.C., 1999, Phys. Lett. A, 256, 347
\bibitem[Jesus(2017)]{jes17} Jesus, J.F., 2017, arXiv:1712.00697 
\bibitem[Kerr(1963)]{ker63} Kerr, R.P., 1963, Phys. Rev. Lett., 11, 237
\bibitem[Kodama(1980)]{kod80} Kodama, H., 1980, Progr. Theor. Phys., 63, 1217
\bibitem[Levinson et al.(2015)]{lev15} Levinson, A., van Putten, M.H.P.M., \& Pick, G., 2015, ApJ, 812, 124
\bibitem[Maeder(2017)]{mae17} Maeder, A., 2017, ApJ, 849, 194
\bibitem[Metzger et al.(2011)]{met11} Metzger, D.B., et al., 2011, Mon. Not. R. Astron. Soc. 413, 2031
\bibitem[{Nakar(2007)}]{nak07} Nakar, E., 2007. Phys. Rep. 442,166
\bibitem[Penrose(1965)]{pen65} Penrose, R., 1965, Phys. Rev. Lett., 14, 57
\bibitem[Perlmutter et al.(1999)]{per99} Perlmutter, S., et al., 1999, ApJ, 517, 565
\bibitem[{Piran(2004)}]{pir04} Piran, T., 2004, RMP, 76, 1143 
\bibitem[Piro \& Pfahl(2007)]{pir07} Piro, A.L., \& Pfahl, E., 2007, ApJ, 658, 1173 
\bibitem[Riess et al.(1998)]{rie98} Riess, A., et al., 1998, ApJ,  116, 1009
\bibitem[Riess et al.(2016)]{rie16} Riess, A.G., Macri, L.M., Hoffmann, S.L., Scolnic, D., Stefano, C., et al., 2016, ApJ, 826, 56
\bibitem[Sari \& Piran(1997)]{sar97} Sari, R., \& Piran, T., 1997, ApJ, 485, 270
\bibitem[Sari et al.(1998)]{sar98} Sari, R., Piran, T., \& Narayan, R., 1998, ApJ, L17
\bibitem[Thornburg(2007)]{tho07} Thornburg, J., 2007, Liv. Rev. Rel., 10, 7 
\bibitem[van Putten \& Levinson(2003)]{van03} van Putten, M.H.P.M., \& Levinson, A., 2003, ApJ, 584, 937
\bibitem[van Putten(2009)]{van09} van Putten, M.H.P.M., 2009, MNRAS, 396, L81
\bibitem[van Putten et al.(2011a)]{van11} van Putten, M.H.P.M., Kanda, N., Tagoshi, H., Tatsumi, D., Masa-Katsu, F., \& Della Valle, M., 2011a, Phys. Rev. D 83, 044046
\bibitem[van Putten(2012a)]{van12a} van Putten, M.H.P.M., 2012a, Prog. Theor. Phys., 127, 331
\bibitem[van Putten et al.(2014a)]{van14a} van Putten, M.H.P.M., Guidorzi, C.., \& Frontera, P., 2014a, ApJ, 786, 146
\bibitem[van Putten et al.(2014b)]{van14b} van Putten, M.H.P.M., Lee, G.M., Della Valle, M., Amati, L., \& Levinson, A., 2014, MNRASL, 444, L58
\bibitem[van Putten(2015)]{van15} van Putten, M.H.P.M. 2015, MNRAS, 450, L48
\bibitem[van Putten(2016)]{van16} van Putten, M.H.P.M., 2016, ApJ,  819, 169
\bibitem[van Putten \& Della Valle(2017)]{van17a} van Putten, M.H.P.M., \& Della Valle, 2017, MNRAS, 464, 3219
\bibitem[van Putten(2017b)]{van17b} van Putten, M.H.P.M., 2017, PTEP, 93F01
\bibitem[van Putten(2017c)]{van17c} van Putten, M.H.P.M., 2017, ApJ, 848, 28
\bibitem[van Putten(2018)]{van18} van Putten, M.H.P.M., 2018, under review
\bibitem[Wald \& Iyer(1991)]{wal91} Wald, R.M., \& Iyer, V., 1991, Phys. Rev. D, 44, R3719
\bibitem[Woosley(1993)]{woo93} Woosley, S.L., 1993, ApJ, 405, 273
\bibitem[Woosley \& Bloom(2006)]{woo06} Woosley, S.E., \& Bloom, J.S., 2006, Ann.Rev.Astron.Astrophys., 44, 507
\bibitem[Woosley(2010)]{woo10} Woosley, S.E., 2010, ApJ, 719, L204
\bibitem[York(1989)]{yor89} York, J.W., 1989, in Frontiers in Numerical Relativity, ed. C.R. Evans, L.S. Finn \& D.W. Hobill 
    (Cambridge University Press)
\bibitem[{Zigao et al.(2017)}]{zig17} Zigao, D., Fr\'ed\'eric, D., \& M\'esz\'aros, P., 2017, SSRv, 212, 409
\bibitem[Zhang et al.(2016)]{zha16} Zhang, B., L\"u, H.-J., \& Liang, E.-W., SSR, 202, 3

\end{thebibliography}

\end{document}